\documentclass[11pt,draft]{asme2ej}

\usepackage{epsfig}
\usepackage{graphicx}
\usepackage{fancyhdr}
\usepackage{setspace}
\usepackage{helvet}
\usepackage[hyphens]{url}
\usepackage[titles]{tocloft}

\addtocontents{lof}{\cftpagenumbersoff{}}
\addtocontents{lot}{\cftpagenumbersoff{table}}
\makeatletter
\renewcommand\@dotsep{400}
\makeatother

\usepackage{amsmath}
\usepackage{physics}
\usepackage{chemmacros}

\usepackage{array}
\usepackage{array,booktabs}
\newcolumntype{P}[1]{>{\centering\arraybackslash}p{#1}}
\newcolumntype{M}[1]{>{\centering\arraybackslash}m{#1}}
\usepackage{multirow}

\usepackage{graphicx}
\usepackage{subfig}
\usepackage{float}
\usepackage[bottom]{footmisc}

\usepackage{soul}

\title{Boltzmann-Curtiss Description for Flows under Translational Nonequilibrium}

\author{Mohamed M. Ahmed
    \affiliation{
        Department of Mechanical and Nuclear Engineering \\
        Kansas State University, Manhattan, KS 66502, USA\\
        }
}

\author{Mohamad I. Cheikh and James Chen \thanks{Address all correspondence for to this author.}
    \affiliation{ 
        Department of Mechanical and Aerospace Engineering \\
        University at Buffalo -- The State University of New York, Buffalo, NY 14260, USA\\
        chenjm@buffalo.edu \\
        }
}

\begin{document}
\maketitle    
\doublespacing
\section*{Abstract}
{\noindent \it Continuum-based theories, such as Navier-Stokes equations, have been considered inappropriate for flows under nonequilibrium conditions. In part, it is due to the lack of rotational degrees of freedom in the Maxwell-Boltzmann distribution.
The Boltzmann-Curtiss formulation describes gases allowing both rotational and translational degrees of freedom and forms morphing continuum theory (MCT).
 The first order solution to Boltzmann-Curtiss equation yield a stress tensor that contains a coupling coefficient that is dependent on the particles number density, the temperature and the total relaxation time. A new \textcolor{black}{bulk viscosity} model derived from 
\textcolor{black}{the Boltzmann-Curtiss distribution is employed} for shock structure and temperature profile under translational and rotational nonequilibrium. Numerical simulations of argon and nitrogen shock profiles are performed in the Mach number range of 1.2 to 9. 
The current study, when comparing with experimental measurements and Direct Simulation Monte Carlo (DSMC) simulation, show a significant improvement in the density profile, normal stresses and shock thickness at nonequilibrium conditions than Navier-Stokes equations. \textcolor{black}{The results indicate that equations derived from the Boltzmann-Curtiss distribution are valid for a wide range of nonequilibrium conditions than those from the Maxwell-Boltzmann distribution. }}

\newpage
\section*{Introduction}

In supersonic and hypersonic flows, shock waves redistribute the high kinetic energy of the flow into different internal energy modes. The characteristic flow length scale inside a shock wave is in the order of the local mean free path of the colliding gas particles. If the number of intermolecular collisions is not sufficient enough, the distribution of energy modes will be relaxed slowly leading to thermal and chemical nonequilibrium inside and behind the shock wave, i.e. near the stagnation point. This can have significant influences on the overall aerodynamic and thermal loadings of aerospace vehicles. Therefore, researchers have been actively studying the shock structure under nonequilibrium conditions, and to find accurate theoretical models that can be implemented in numerical simulations \cite{Hejranfar_2019}.

While energy is exchanged during intermolecular collisions, it is assumed that the distribution of molecular energies among various energy modes is characterized by a Maxwell-Boltzmann distribution. This distribution should not vary with time at a thermal equilibrium state. If the gas molecules undergo chemical reactions, chemical equilibrium also implies that the composition of the gas should not change with respect to time. Therefore, a certain number of molecular collisions must take place in order for each energy mode to reach equilibrium, i.e. a Maxwellian energy distribution \cite{vincenti1965introduction}. The energy modes that are typically considered in aerothermodynamics are those associated with translation, rotation, vibration and electronic excitation of molecules. The translational degree of freedom is described in the framework of classical mechanics. The other internal energy modes of rotation, vibration and electronic are quantized, and usually described by quantum physics. In monatomic gases, translational nonequilibrium may exist if the molecules did not undergo sufficient collisions, and only electronic excitation may occur during intermolecular collisions. On the other hand, for diatomic and polyatomic gases, in addition to translational and electronic energy, rotational and vibrational energy modes arise in light of the interaction of the two atoms around their chemical bond. Rotations and vibrations are independent to one another, and to translational motion as well. 

As the Mach number increases beyond sonic limit, the distribution of molecular velocities deviates from the \textcolor{black}{classical} Maxwellian distribution. Since Navier-Stokes (NS) equations are derived \textcolor{black}{from} Chapman-Enskog method assuming slight departure from translational equilibrium, its capability to capture the translational and rotational nonequilibrium physics appears to be limited. Consequently, it was found when comparing the density profiles obtained from the NS simulations with experimental data that the shock thickness is much thinner in NS solutions \cite{BoydChenCandeler, schwartzentruber2006hybrid}. On the other hand, the DSMC method provides a better description for the shock wave structure under nonequilibrium conditions since it encounters the intermolecular interactions and the effect on the overall thermal equilibrium of the flow. Several techniques have developed for DSMC \cite{Shah_2018}. However, the DSMC method can not be applied to flows at high densities due to the expensive numerical cost.

In the classical kinetic theory, volumeless point particles are considered for describing the behavior of a gas at microscopic level. Due to the difficulty of tracking particle properties before and after it collides with another particle, a statistical approach of introducing a probability distribution function of molecular velocities is introduced. This approach leads to the classical integro-differential Boltzmann equation. Two main \textcolor{black}{linearizing} approaches were introduced in order to solve the Boltzmann equation. The first approach, which is provided by Bhatnagar, Gross and Krook (BGK), described essential molecular interactions. However, it neglected the details of the nonlinear collision integral in Boltzmann equation \cite{BGK}. A more accurate approach was obtained by Chapman and Enskog (CE). In the CE approach, a power series expansion about the equilibrium distribution function was considered. The solution of the Boltzmann equation considering only the first order term of the CE power series, i.e. slight deviation from equilibrium, \textcolor{black}{leads} to Navier-Stokes equations \cite{vincenti1965introduction,boyd2017nonequilibrium}. Also, Burnett and super Burnett equations can be obtained if second order and higher order terms are considered. Although these equations can account for larger deviation from equilibrium, they are subject to numerical instabilities. It should be noted that only one viscosity coefficient were obtained from the kinetic theory representation of the Navier-Stokes equations, which is the shear viscosity $\mu$.

Navier-Stokes equations can also be deduced under the framework of rational continuum mechanics by requring objective (frame-indifferent) constitutive equations and no body couples act on the fluid surface, i.e. symmetric stress tensor \cite{truesdell2010introduction,eringen1980mechanics}. These two axioms reduce the material parameters, that relate the stress acting on the fluid surface to its deformation, to only two, namely the shear viscosity $\mu$ and the second viscosity coefficient $\lambda'$. It is assumed in most fluid applications, following the Stokes hypothesis, that the dilatation effects can be neglected and the bulk viscosity ($\mu_B=\lambda'+\frac{2}{3}\mu$) can be set to zero, i.e. $\lambda'=-\frac{2}{3} \mu$ \cite{graves1999bulk}. Thus, only one viscosity coefficient ($\mu$) can be used to describe the fluid flow. Although this assumption is supported by the kinetic theory for the Navier-Stokes equations \cite{vincenti1965introduction}, it may not be valid for monatomic gases that deviate from translational equilibrium, or for diatomic gases under rotational nonequilibrium. For instantance, several studies on shock wave structure have shown that shock profiles obtained from the solution of Navier-Stokes equations deviates from Direct simulation Mote Carlo solutions (DSMC) and experimental data at strong hypersonic conditions \cite{BoydChenCandeler, continuum_breakdown1,continuum_breakdown2}, as well as the solutions from the higher order Burnett and super Burnett equations \cite{fiscko1989comparison}.

Since the rotational relaxation time at moderate temperatures is in the same order, or slightly greater than the translational relaxation time, it is argued that the rotational and translational energies are assumed to be in equilibrium with one another, and the effect of a weak departure from either translational or rotational equilibrium can be accounted for if the bulk viscosity is retained as an adjustable parameter in the dissipative terms in the NS equations \cite{vincenti1965introduction}. However, to the best of the authors knowledge, no rigorous derivation is found in the literature to relate the bulk viscosity with either translational or rotational nonequilibrium. Although, this assumption contradicts the classical kinetic theory representation of the Navier-Stokes equations, in which zero bulk viscosity were obtained, experimental investigations show that the bulk viscosity values of dense and dilute gases can be in the order of, or exceed, the shear viscosity \cite{madigosky1967density}.

Most experimental effort to interpret the true values of the bulk viscosity of dense and dilute \textcolor{black}{gaseous} fluids involved acoustic absorption measurement of sound waves. \textcolor{black}{Majority} of the available experimental results in the literature are summarized by Graves at al. \cite{graves1999bulk}. For instance, some data reported a ratio between the bulk viscosity to shear viscosity of order of 2/3 for air and 2000 for CO$_2$ and N$_2$O. On the other hand, the assumption of the bulk viscosity as an adjustable parameter to account for strong deviations from translational and rotational nonequilibrium has been employed in different studies on shock wave profiles. Chapman et al. introduced models of bulk viscosity to Navier-Stokes equations that yield more accurate shock-wave density profiles but failed to accurately predict temperature profiles. They also introduced nonlinear stress and heat flux tensors that produce considerably more realistic results than Navier-Stokes equations \cite{ChapmanShock}. 

Alternative theoretical approaches to Navier-Stokes theory for modeling nonequilibrium gas flows emerged from statistical mechanics and rational continuum mechanics. Firstly, Dahler \cite{dahler1} extended statistical theories of transport phenomena to fluids with translational and rotational degrees of freedom. \textcolor{black}{Dahler} derived the general hydrodynamic equations that describe a dilute fluid composed of diatomic molecules. In addition to presenting the familiar conservation equations of mass, linear momentum and energy, he presented a new equation associated with the conservation of angular momentum. Few years later, Dahler and Scriven \cite{dahler2} presented a similar change of internal angular momentum equation for a continuum model of matter. They concluded that the system may exchange external angular momentum with the surrounding or accumulate it within the system and this process results to an asymmetrical state of stress. This asymmetry originates from the interaction of the center of mass system with an inter-penetrating twist or spin system, i.e. external body couples, which is common during the collision of diatomic or polyatomic molecules. \textcolor{black}{Taniguchi et al. developed a theory based on extended thermodynamics which was found to accurately describe the density profiles of $CO_2$ shock waves \cite{taniguchi2014thermodynamic}. Kosuge et al. also investigated the shock structure of of $CO_2$ gas using a simple model Boltzmann equation \cite{kosuge2016shock}. Xu and Josyula provided a generalized extension to NS equations based on the Bhatnagar-Gross-Krook (BGK) model of Boltzmann equation \cite{xu2006continuum}. In addition, several other efforts have also been introduced for polyatomic molecules \cite{Boltzmann_1964,Wu2015,Bisi2017}. Nevertheless, these studies mostly focus on the internal mode of rotation, but omit macroscopic spining motion. Recently, Chen and colleagues form a morphing continuum theory (MCT) for high speed aerodynamics \cite{ChenAKT, wonnell2018first, chen2017morphing, PRF024604, ChenAIAA2018} by expanding the Curtiss' kinetic formulations for polyatomic molecules and its local spinning motion \cite{curtiss2,curtiss3}.}

In this article, the Boltzmann-Curtiss distribution for morphing continuum is considered for investigating the shock structure using a first order approximate solution to Boltzmann-Curtiss equation. The first section provides a mathematical and physical description of the formulation. In section 2, a discussion on the stress tensor obtained from this solution and its relation to thermal nonequilibrium is conducted, and followed by an interpretation of the material parameter, which is equivalent to the adjustable bulk viscosity in NS equations. The results of the numerical simulations of argon and nitrogen shock structures are presented and discussed in section 3. Finally, a brief discussion about the current study and future perspectives is provided in the last section. 

\section*{The Boltzmann-Curtiss equation}
\textcolor{black}{ The starting point for governing the transport phenomena of low-density, nonreacting gases is the classical Boltzmann equation. The Boltzmann equation assumes gas molecules as point particles based on many physical arguments, and tracks only their position and velocity. The current work, however, is based on the extension of the Boltzmann transport equation by Curtiss \cite{curtiss3}. Curtiss describes the molecules not as point particles, but as finite size structures with additional variables associated with their internal vibration, and internal spin, apart from the translational velocity \cite{curtiss1, curtiss2, curtiss3}. If the internal vibration and external forces are omitted, the Boltzmann-Curtiss kinetic equation for diatomic (and linear polyatomic) molecules can be expressed \cite{curtiss3, wonnell2018kinetic} as the following:
}
\begin{equation}
\textcolor{black}{
\left(\frac{\partial}{\partial t} + \frac{p_i}{m} \frac{\partial}{\partial x_i} + \frac{M_i}{I} \frac{\partial}{\partial \Phi_i}\right)f(x_i,v_i,\omega_i,\Phi_i, t)=R[f]
}
 \label{eqn0} 
\end{equation}
\textcolor{black}{
where $f$, $m$, $x_i$, $p_i$, $M_i$, $\Phi_i$ ,and $I$ represent the probability distribution function, the mass of the particle, the particle's position, the particle's linear momentum, the particle's angular momentum, the Euler angle associated with the orientation of the particle, and the particle's moment of inertia, respectively. As for $R[f]$ it represents the Boltzmann-Curtiss collision integral. 
\\
The distribution function $f(x_i,v_i,\omega_i,\Phi_i, t)$ gives the probability measure a particle possess in a system over various possible states. For instance, the current distribution function describes the location, speed, gyration, and Euler angle a particular particle owns at a given time. The function is absent of dependencies on vibrational energy or vibrational motion, thus the dynamics of the collision integral is assumed to be independent of these variables.
\\
 For the current work, the particles are treated as spheres, so all dependencies on the axial orientations are omitted, i.e. dependency on the Euler angle "$\Phi_i$" drops . Therefore, the Boltzmann-Curtiss kinetic equation becomes}
\begin{equation}
\textcolor{black}{
\left(\frac{\partial}{\partial t} + \frac{p_i}{m} \frac{\partial}{\partial x_i}\right)f(x_i,v_i,\omega_i,t)= R^*[f]
}
\label{eqn1} 
\end{equation}
\textcolor{black}{The right hand side of this equation accounts for cumulative effect of particles collisions on the distribution void of any axial orientations.}

 Since the remaining terms are concerned with linear momentum, \textcolor{black}{one would expect} the equilibrium solution, or the zeroth-order solution, to this equation to be similar to the Maxwell-Boltzmann distribution function. However, the existence of an independent degree of freedom of gyration $\omega_i$ changes the distribution of particles kinetic energy. The equilibrium distribution function that \textcolor{black}{approximates the solution to} equation \ref{eqn1} can be obtained from Boltzmann's principle \cite{ChenAKT}. \textcolor{black}{A zeroth-order approximation to the} equilibrium distribution function is \cite{ChenAKT, wonnell2018first}
\begin{equation}
\begin{split}
f^0(v'_i,\omega'_i&,x_i,t)=\\
n&\left(\frac{mI}{4\pi^2 \theta^2}\right)^{3/2}  \exp(-\frac{m v'_l v'_l+I \omega'_p \omega'_p}{2\theta})
\end{split}
 \label{eqn2}
\end{equation}
where $v'_l=v_l-U_l$ is the perturbed velocity from the mean velocity $U_l$, $\omega'_l=\omega_l-W_l$ is the perturbed gyration from the mean gyration $W_l$, $\theta=kT$ is the thermal energy, $k$ is Boltzmann's constant, $T$ is the temperature, $n$ is the number density, and $I$ is the moment of inertia of the spherical particle. The moment of inertia can be approximated as $I=m j$, where $j=\frac{2}{5} r^2$, and $r$ is the spherical particle \textcolor{black}{radius} \cite{chen2012numerical}. \textcolor{black}{The Boltzmann-Curtiss distribution differs from the classical Boltzmann distribution function by including an additional contribution to the momentum from the perturbed gyration, $\omega_p'$. The number density, $n$, of the particles is found by integrating the distribution function $f^0$ over all the perturbed variables,}
\begin{equation}
n= \iint f^0 \ dv_i' \ d\omega_i' \label{eqn3}
\end{equation}
\textcolor{black}{Derivation of the Boltzmann-Curtiss distribution can be found in the author's previous publication \cite{ChenAKT} }. Figure \ref{fig1} shows a graphical representation of the \textcolor{black}{Boltzmann-Curtiss} distribution function \textcolor{black}{for a one-dimensional flow. The highest point in a Boltzmann-Curtiss distribution function is referred to as the most probable velocity and gyration the largest number of molecules can have.} 

It can be seen that the Maxwell-Boltzmann distribution can be recovered by projecting the Boltzmann-Curtiss distribution to any plane of a constant ${\omega_p}$. In other words, the Boltzmann-Curtiss distribution resembles a higher order domain for equilibrium than the one represented by classical Maxwell-Boltzmann distribution. Therefore, not only the equilibrium state represented by the Maxwell-Boltzmann distribution can be described, but also any deviation from such distribution for nonequilibrium states can be captured.

\subsection*{First Order Approximation}
\textcolor{black}{The equilibrium distribution function $f^0$, provides an abstract description for the kinematics of diatomic (and linear polyatomic) molecules. The function shows that any conserved quantity ($\chi$) in the system has to be associated with the velocity of the molecule $v_i$, as well as its gyration $\omega_i$. Therefore, balance laws governing any conserved quantity $\chi$, must arise by averaging the Boltzmann-Curtiss transport equation \ref{eqn1}. At equilibrium, and in the absence of collision the transport equation for $\chi$ becomes \cite{ChenAKT, wonnell2018first}:
\begin{equation}
\pdv{\langle n \chi \rangle}{t}+\frac{\partial}{\partial x_i}\langle n \frac{p_i}{m} \chi \rangle - n \langle \frac{p_i}{m} \pdv{\chi}{x_i} \rangle =0
\end{equation}
where $\langle \cdot \rangle$ is the average of the said quantity defined by the following definition:
\begin{equation}
\langle A \rangle= \frac{1}{n} \iint A f(x_i,v_i,\omega_i,t) d^3 v' d^3 \omega'
\end{equation}
and $n$ is the number density of the particles found by integrating the distribution
function $f$ over all the perturbed variables, $v'$ and $\omega'$. } Substituting the conserved quantities of mass, linear momentum, angular momentum and energy for $\chi$, letting the average of the velocity $\vec{v}$ and gyration $\vec{\omega}$ equal to their mean and splitting total variables into mean and fluctuating components, the balance laws become \cite{ChenAKT,wonnell2018kinetic}:

\textbf{Continuity $(\chi_1=m)$}
\begin{equation}
\pdv {\rho}{t}+\frac{\partial}{\partial x_l} (\rho U_l)=0
\label{Eq_cont}
\end{equation}

\textbf{Linear Momentum $(\chi_2=m(v_i+\epsilon_{ipl}r_l\omega_p))$}
\begin{equation}
\pdv{\rho U_s}{t} + \frac{\partial}{\partial x_l}(\rho U_s U_l)- \frac{\partial}{\partial x_l}t_{ls}=0
\label{Eq_LM}
\end{equation}

\textbf{Angular Momentum $(\chi_3=m(r_i r_p \omega_p))$}
\begin{equation}
\frac{\partial}{\partial t}(\rho j W_s) + \frac{\partial}{\partial x_l}(\rho j W_s U_l)- \frac{2}{3} \frac{\partial}{\partial x_l}(m_{ls})=0
\label{Eq_AM}
\end{equation}

\textbf{Energy $(\chi_4=m(e+\frac{1}{2}[v'_l v'_l+r_p r_q \omega'_p \omega'_q]))$}
\begin{equation}
\pdv{\rho e}{t}+\frac{\partial}{\partial x_l}(\rho e U_l) +\pdv{q_l}{x_l}- \rho \langle v_l \pdv{e}{x_l} \rangle=0
\label{Eq_E}
\end{equation}
where, the stress \textcolor{black}{tensor, moment tensor} and heat flux \textcolor{black}{vector} are expressed as:
\begin{align}
t_{kl}&=-\rho \langle v'_k v'_l \rangle -\rho \langle v'_k \epsilon_{lpq} r_q \omega'_p \rangle \label{tkl}\\
m_{kl}&=-\rho \langle \frac{3j \omega'_k v'_l}{2}\rangle \label{mkl} \\
q_k&=\frac{1}{2} \langle \rho v'_l v'_l v'_k + \frac{3j \omega'_p \omega'_p v'_k}{2}\rangle \label{qk}
\end{align}

\textcolor{black}{Chen \cite{ChenAKT} employed the zeroth-order approximation to the equilibrium distribution to derive a form for the stresses and heat flux. The resulting balance laws showed similarity to the Euler equations with an additional law governing the angular momentum.}

\textcolor{black}{The current study follows the work by Wonnell and Chen \cite{wonnell2018kinetic, wonnell2018first} which is based upon a higher order approximation for the distribution function.} For an equilibrium distribution function, a large number of binary collisions is assumed to occur in a short period of time. In other words, any deviation from an initial state of equilibrium distribution would return rapidly back to final state of equilibrium. These binary collisions affect the initial and final velocities and gyrations of the particles. Since gyration is treated as a classical motion applicable to the same molecules, similar to the velocity, the collision integral on the right-hand side of Boltzmann-Curtiss transport equation can be obtained from Wang Chang-Uhlenbeck description \cite{wang1964heat}, with an additional term to the linear momenta pre and post collision, which accounts for the contribution from the component of local rotation moving in the direction of translational velocity. \textcolor{black}{For this reason the collision integral $R^*[f]$ is given by the following integral \cite{wonnell2018kinetic, wonnell2018first}:
\begin{equation}
\begin{split}
R^*[f] = \iiint d^3 p_2  d^3 p_1' d^3 p_2'& \delta^4(P_f - P_i)\\
&|T_{fi}|^2 (f_2'f_1'-f_2 f_1)
\end{split}
\label{eqn41}
\end{equation}
where $P_i$ and $P_f$ are the initial and final total momenta, $p_1$ and $p_2$ are the initial momenta of the colliding particles, and their primed counter parts are their respective final momentum. The current Boltzmann-Curtiss linear momentum differs from the classical linear momentum $p_i = m v_i$, by including an additional contribution from the component of the local rotation moving in the direction of the translational velocity \cite{cassidy2004},
\begin{equation}
p_i=m(v_i+\epsilon_{ipl} r_l \omega_p) \label{eqn4}
\end{equation}
}where $r_l$ is the distance vector emerging from the center of mass of the particle and $\epsilon_{ipl}$ is the Levi-Civita tensor. \textcolor{black}{ The matrix $T_{f_i}$ converts the particle from its pre collision state to its post collision state.} If the transition from initial to final state is assumed to consume a time $\tau$, a first order estimate to the collision integral can be expressed as \cite{wonnell2018kinetic, Huang1987}
\begin{equation}
R^*[f]=-\frac{f-f^0}{\tau}=-\frac{g}{\tau} \label{eqn5}
\end{equation}
where $g=f-f^0$. \textcolor{black}{The first-order distribution function, $g$, measures the probability that the molecules will exit their equilibrium state through collisions only. As for $\tau$, it characterizes  the transition of the velocity and gyration to and from the equilibrium state.} Substituting equation \ref{eqn5} into the equation \ref{eqn1} yields an approximate form for the Bhatnagar, Gross, and Krook (BGK) equation \cite{BGK}:
\begin{equation}
g=-\tau \left(\frac{\partial}{\partial t}+v_i \frac{\partial}{\partial x_i} \right)(f^0+g)
\end{equation} \label{eqn6}
Since \textcolor{black}{the study is } concerned only with the forces and properties influencing the mean flow when a slight deviation from equilibrium occurs, it is assumed that $g << f^0$. Hence, a simplified formula for finding $g$ from the derivatives of $f^0$ can be obtained:
\begin{equation}
g=-\tau\left(\frac{\partial}{\partial t}+v_i \frac{\partial}{\partial x_i}\right)f^0
\end{equation} \label{eqn7}
Using chain rule \textcolor{black}{on the spatial derivatives of $f^0$}, and after some mathematical manipulation, an expression of $g$ is obtained \cite{wonnell2018kinetic, wonnell2018first}:
\begin{equation}
\begin{split}
g=&-\tau f^0 \left[\frac{1}{\rho}\left(v'_i \pdv{\rho}{x_i}-\rho \pdv{U_i}{x_i}\right) \right.\\ 
&-\left(\frac{3}{\theta}-\frac{m(v'_l v'_l+j \omega'_l \omega'_l)}{2\theta^2}\right)\left(v'_i \pdv{\theta}{x_i}-\frac{\theta}{3} \pdv{U_q}{x_q}\right) \\ 
&\left. +\frac{m v_i}{\theta}\left(v'_l \pdv{U_i}{x_l} - \frac{1}{\rho} \pdv{n \theta}{x_i}\right) +\frac{m j \omega'_i}{\theta}\left(v'_l \frac{W_i}{x_l}\right)\right]
\end{split}
  \label{eqn8}
\end{equation}
\textcolor{black}{ Here, $g$, represents a first order deviation from the equilibrium distribution function $f^0$ expressed entirely in terms of the mean and perturbed flow properties. Substituting $g$ in equations \ref{tkl}-\ref{qk}}, a first order approximation to the stress \textcolor{black}{tensor, moment tensor} and heat flux  can be obtained \textcolor{black}{\cite{wonnell2018first}} and recover the constitutive equations for morphing continuum theory (MCT) \cite{chen2017morphing, PRF024604, ChenAIAA2018}:
\begin{align}
\begin{split}
t_{kl}=& -p \delta_{kl} - \frac{\eta}{3} \pdv{U_m}{x_m} \delta_{kl}  +\eta\left(\pdv{U_l}{x_k}+\pdv{U_k}{x_l}\right) \\
& \textcolor{black}{+ \eta \epsilon_{lmn} r_n \pdv{W_m}{x_k} }
\label{stress}
\end{split}
\\
m_{kl}&= \eta  \frac{3j}{2} \pdv{W_l}{x_k}
\label{mom-stress}
\\
q_k &= -4 \frac{\eta}{m} \pdv{\theta}{x_k}
\label{heat-flux}
\end{align}
where
\begin{equation}
   \eta=n \tau \theta \quad , \quad \theta=k T
\end{equation}

The next step is to find an expression for the material parameter $\eta$ for numerical simulations of shock wave profiles at hypersonic Mach numbers.


\section*{A Boltzmann-Curtiss Bulk Viscosity Model}

In the classical continuum theory, the fluid element represents a group of material points that are not oriented, and the rotational motion of the fluid can only be deduced from the translational motion of the material points. Therefore, no additional equation of rotation is considered and no coupling terms of rotational motion appear in the stress tensor. The stress tensor in NS equations is defined by
\begin{equation}
t_{kl}^{NS}=-p \delta_{kl} + \lambda' \pdv{U_m}{x_m} \delta_{kl}+ \mu \left( \pdv{U_l}{x_k}+\pdv{U_k}{x_l} \right)
\end{equation}
Following Stokes' hypothesis, if the kinetic pressure \textcolor{black}{$p^*=n k \theta$} is assumed to be equivalent to only the thermodynamic pressure $p$, and the contribution of fluid velocity on the normal stresses \textcolor{black}{are} neglected, i.e. $-1/3 t_{kk}=p$, the following relation should be satisfied
\begin{equation}
\lambda'=-\frac{2}{3} \mu
\end{equation}
The gas kinetic representation of NS equations, obtained from Chapman-Enskog solution of the classical Boltzmann equation of point particles, supports this assumption, and the stress tensor obtained takes the form
\begin{equation}
t_{kl}^{NS}=-p \delta_{kl} -\frac{2}{3} \mu \pdv{U_m}{x_m} \delta_{kl}+ \mu \left( \pdv{U_l}{x_k}+\pdv{U_k}{x_l} \right)
\end{equation}

As for MCT, the $r_n$ in the last term ($\eta \epsilon_{lmn} r_n \frac{\partial W_m}{\partial x_k}$) of the MCT stress indicates the length scale of the local rotation and is often insignificant while comparing with the characteristic length in the flows, i.e.  

\begin{equation}
t_{kl}^{MCT}= -p \delta_{kl} - \frac{\eta}{3} \pdv{U_m}{x_m} \delta_{kl}  +\eta\left(\pdv{U_l}{x_k}+\pdv{U_k}{x_l}\right)
\end{equation}

It leads to a new relation for bulk viscosity
\begin{equation}
\eta = \mu \qquad\text{and}\qquad\lambda'=-\frac{1}{3}\mu
\end{equation}

Unlike the bulk viscosity term added to the NS stress tensor, the term $\frac{\eta}{3} \pdv{U_k}{x_k}$, where $\eta=n \tau \theta$, is inherent in the stress tensor obtained from the first order solution to Boltzmann-Curtiss equation. It can automatically account for any deviations from both translational and rotational equilibrium, since the relaxation time here is the time required for the system to reach full thermal equilibrium, i.e. not only translational equilibrium but equilibrium in both rotational and translational motions.

\section*{Nonequilibrum shock wave structure}
The current section employs the first order approximation to the Boltzmann-Curtiss equation to investigate the shock structure in monatomic and diatomic gases. The problem of shock wave structure is selected as it represents a flow regime which is far from thermodynamic equilibrium. Moreover, this problem deals with a one-dimensional flow, in which the impact of the mean gyration of the flow is null, and the gyration equation behaves as transport equation carrying a flow property that doesn't contribute to the physics of the flow. Also, the complexity of the boundary condition selection and its interference with the solution is eliminated by considering this problem. Therefore, this problem focuses on the effect of the bulk viscosity model for the shock wave profiles.

By employing the previous assumptions, and recovering the definition of $\eta$, the stress tensor  becomes symmetric, i.e. 
\begin{align}
t_{kl} &= -p \delta_{kl} - \frac{\mu}{3} \pdv{U_m}{x_m} \delta_{kl} + \mu \left( \pdv{U_l}{x_k} + \pdv{U_k}{x_l} \right)
\label{MCTStress}
\end{align}

With the new interpretation of the material parameter $\eta$, the MCT heat flux vector in equation \ref{heat-flux} becomes

\begin{equation}
    q_k = -4 \mu \frac{k}{m} \pdv{T}{x_k} \label{thermal_cond}
\end{equation}

This heat flux exhibits a linear relation with temperature gradient, similar to Fourier's law of thermal conduction, with a thermal conductivity dependent on the product of the temperate and the relaxation time. However, the new expression for the thermal conductivity 
shows that the specific heat is approximately four times the gas constant if a unity Prandtl number is considered. This ratio agrees with the estimated range reported in the NASA technical report \cite{svehla1962estimated} for the calculation of the viscosity and thermal conductivity of gases based on Lennard-Jones potential, which shows that the ratio for diatomic nitrogen is varying between $3.501$ to $4.545$ at ranges of temperature from $100-5000 K$.

With new expressions for the stress tensor and heat flux vector, they can be plugged back into the balance laws described earlier for validation with the DSMC results and experimental measurements and direct comparison with the nonequilibrium solution from NS equations. 
A power law model of the relaxation time as a function of the temperature is employed. Note that this approximation does not physically contradict with the derivation of the material parameter $\mu$, which is a function of the product of the temperature and the relaxation time. It, instead, ensures a consistent intermolecular interaction model while comparing with DSMC and NS solutions (e.g. \cite{schwartzentruber2006hybrid,BoydChenCandeler}). 

The specific heats and the internal energy of the molecules are obtained in accordance with the principle of equipartition of energy. Since a single temperature is used in the proposed formulation to describe thermodynamic equilibrium (i.e. both rotational and translational equilibrium), the Landau-Teller-Jeans equation \cite{jeans} is included in diatomic gas simulations to extract rotational and translational temperatures. The approximate expression for the rotational collision number obtained by Parker \cite{parker1959} is considered for the calculation of the rotational relaxation time constant.

The finite volume solver used in this study employs a forward Euler temporal discretization for the unsteady term. In order to have a nonoscillating solution in a hypersonic flow regime, a second order flux splitting scheme introduced by Kurganov, Noelle and Petrova (KNP) \cite{KNP, PRF024604} is used in the calculation of the convective term at the interfaces. All diffusion terms are computed using central differencing. The spatial accurance has proven to be second order by following a rigorous verification and validation procedure proposed by Roy et. al. \cite{roy1, roy2}. The full description of the code, as well as the verification and validation results, were presented and published in the 2019 AIAA SciTech Forum \cite{ahmed2019verification}.

\subsection*{Shock Structure for Monatomic Gases: Argon}

Numerical simulations of shock wave structure in argon gas are performed at different Mach numbers up to Mach 9 on 3000 uniform cells, which span approximately 40 upstream mean free paths. Any computed flow property $Q$ is normalized in the familiar way
\begin{equation}
Q_{\textrm{normalized}}=\frac{Q-Q_1}{Q_2-Q_1}
\end{equation}
where subscripts 1 and 2 refer to the upstream and downstream of the shock respectively. The axial distance $x$ is normalized by the upstream mean free path $\lambda$.

In order to show an overall comparison between the solution obtained from the current study and the experimental measurements as well as other numerical simulations of the shock structure, the reciprocal shock thickness, $\frac{d \rho}{d x/\lambda}$ at $\rho_{\textrm{normalized}}=0.5$, is calculated and compiled into one graph for the Mach number range from Mach 1.2 to Mach 9. DSMC and NS data are re-printed from the existing literature \cite{schwartzentruber2006hybrid}. As shown in figure \ref{fig2}, NS simulations predict far too thin shock wave, while the DSMC solution perfectly agrees with the experimental data. The current solution shows significantly improved shock thickness than NS solution. Also, the current solution and Burnett results, obtained by Fiscko and Chapman \cite{fiscko1989comparison}, are almost identical, and both predict a slightly thinner shock wave than DSMC and experimental results of Alsmeyer \cite{alsmeyer1976density}. It should be noted that at transonic Mach numbers, i.e. the Mach range from 1 to 1.3, all simulations predict the same shock thickness as the experimental data, since the flow is still at thermal equilibrium. However, as the Mach number increases to supersonic and hypersonic speeds, the shock thickness predicted by NS simulations significantly deviates from DMSC and experimental data. The current study agrees well with the experimental measurements with only less than 10 \% difference while the NS simulation results show as much as $60 \%$ differences.

A more detailed representation of the density profile at Mach 3.38 is shown in figure \ref{fig3}. The reciprocal shock thickness, $\frac{d \rho}{d x/\lambda}$, confirms that the first order solution to the Boltzmann-Curtiss equation predicts a more realistic thicker shock wave and closer to experimental and DSMC predictions than the nonequilibrium NS solution. This improvement can be explained as following: equilibrium in Boltzmann distribution function is restricted to \textcolor{black}{translational} motion only. However, equilibrium in Boltzmann-Curtiss theory requires not only velocity, but both gyration and velocity of molecules dependently to obey the velocity-gyration based Boltzmann-Curtiss distribution.  When the gyration is zero, i.e. for moantomic gases, the velocity distribution function recovers to the classical Boltzmann velocity distribution function. The nonequilibrium distribution function, which lead to NS equations, assumes a first order deviation from translational equilibrium only (i.e. equilibrium velocity distribution not velocity-gyration distribution). However, the nonequilibrium distribution function obtained from the first order solution to Boltzmann-Curtiss equation extends nonequilibrium to rotational motion. Since only few collisions are required for translational equilibrium, the time required to recover rotational equilibrium is greater than that for translational equilibrium. Hence, the relaxation time employed in the first order approximation to Boltzmann-Curtiss equation in the current study is larger when compared to the relaxation time in the first order approximation to Boltzmann equation that lead to NS equations. Therefore, the nonequilibrium distribution function obtained in the current study represents further departure from translational equilibrium than the classical Maxwellian nonequilibrium distribution function for monatomic gases. In other words, the bulk viscosity of the stress tensor derived from the Boltzmann-Curtiss distribution is accounting for this deviation.

Figure 
\ref{Argon_stress_q}(a-b) shows 
 the normal stress distribution obtained from the currents solution, NS solution and DSMC results for Mach 1.2 and 8. In order to obtain a comparison with DSMC results at a wide Mach number range, the DSMC data are printed from different sources \cite{josyula2004internal,bird1970aspects}. The stresses in $x$ and $y$ directions are normalized in a similar way as described in \cite{josyula2004internal,bird1970aspects}, where $\p1$, $\rho_1$ and $C_1$ are the upstream pressure, density and most probable molecular velocity respectively. The corresponding heat flux is shown in figure
 \ref{Argon_stress_q}(c-d).

At Mach 1.2, NS solution predicts slightly higher peak stresses and much higher heat flux than DSMC solution, while the current solution exactly matches DSMS results. At Mach 8, NS solution significantly deviates further from DSMC results in both stress and heat flux, while the current solution still shows an overall improvement in the prediction of the normal stresses and heat flux as anticipated.

\subsection*{Shock Structure for Diatomic Gases: Nitrogen}

The next set of simulations is carried out for nitrogen, a diatomic gas. Solutions are generated at Mach numbers of 1.2, 5 and 10 for upstream conditions of $\rho_1=1.225 \ kg/m^3$ and $T_1=300 \ K$. The overall reciprocal shock thicknesses is compared to experimental data of Alsmeyer \cite{alsmeyer1976density}, Camac \cite{camac1964argon}, NS and DSMC results \cite{BoydChenCandeler} as shown in figure \ref{nitrogen_thickness}. Similar to the argon case, the current solution successfully predicts a wider shock and a more accurate density profiles than NS solution at hypersonic conditions where both translational and rotational nonequilibrium exist. Also, the current study is shown to agree well with the DSMC solution.

More detailed comparisons of the density and temperature profiles are shown in 
figure \ref{N2_All}(a-f)
for Mach numbers 1.2, 5 and 10. At Mach 1.2, the density and temperature profiles of the three solutions are almost identical with a slight difference between the translational (black) and rotational (blue) temperatures. As the Mach number increases, NS density and temperature profiles deviate further from DSMC results. Note that DSMC results are taken as a reference as it best matches experimental measurements of density profiles. Also, it is useful for comparing temperature profiles as no experimental data exist in the literature on the temperature profiles inside the shock-wave. Overall, the density profiles obtained from the current solution matches the DSMC predicted profile with noticeable discrepancy for the temperature profile. However, since there is a lack of experimental measurements of temperature inside the shock wave, this discrepancy in the temperature profiles requires further investigations. It should also be noted that DSMC is a solution to Boltzmann equation not to Boltzmann-Curtiss equation. 

\section*{Conclusions}
Navier-Stokes equations breaks down when large deviation from thermal equilibrium exist, for example inside a shock wave. Several arguments point out that the bulk viscosity can be a correction parameter added to the NS equations to account for deviations from thermal equilibrium. These arguments, however, are not supported by the kinetic description of the Navier-Stokes equations, i.e. the first order approximation to Boltzmann distribution. 

A first order approximation to the Boltzmann-Curtiss equation, in which gases are represented by spherical particles with rotational and translational degrees of freedom, lead to a more general set of morphing continuum theory equations. Also, the effect of the particle translation and rotation on the stresses in the flow are accounted for through a coupling material parameter in the stress tensor. Further examination of the stress tensor revealed a relationship between this material parameter and the bulk viscosity term in Navier-Stokes equation. Approximate estimation yields that the ratio of the bulk viscosity to shear viscosity is $\frac{1}{3}$.

Numerical simulations of argon shock structure in the Mach number range of 1.2 to 9, which represent transnational nonequilibrium, reveal that the new estimated parameter lead to accurate predictions of the density profiles and shock thickness, when compared to experimental data and DSMC simulations, while NS simulations fail to predict accurate profiles. It is also found that the overall shock thickness predicted in the current study is approximately matching the results of the more complex Burnett equation. The normal stress and heat flux of argon were found to better match the DSMC results than NS at the Mach numbers of 1.2 and 8. These improvements reveal that the nonequilibrium distribution function derived in the current study has extended deviation from equilibrium than the one employed in the derivation of NS equations. Numerical simulations of nitrogen shock structure at translaitonal and rotational nonequilibrium reveal that the current solution accurately predicts the density profiles when compared to available experimental data. Upon the availability of the experimental data, the accuracy of the temperature shall be further evaluated.

\textcolor{black}{Computational resources are always a concern for numerically simulating compressible flows and high speed aerodynamics. In general, continuum-based methods, e.g. NS equations, the presented MCT framework, Burnett equations and others, are more preferred than particle-based approaches, e.g. DSMC and molecular dynamics. However, most continuum methods are either invalid, e.g. NS equations, or inpractical, e.g. Burnett equations. MCT has been shown its advantage in computational resources over NS equations in a few compressible flow cases.  In a case of supersonic turbulence over a compression ramp \cite{PRF024604}, the results indicates that in the wall-normal direction, the smallest grid size near the wall ($\Delta y^+$) is 1.34 (NS: 0.2 in a similar study) with 10 grid points within the critical region ($y^+ < 30$) (NS: 20+ grid points within $y^+ < 20$ in a similar study). In addition, in a case of transonic turbulence over an axisymmetric hill \cite{ChenAIAA2018}, the mesh number for MCT-based DNS ($\sim$6M) is almost an order less than that in NS-based DNS ($\sim$54M). It is worth mentioning even with a coarser mesh, the proposed MCT simulation captures both shocks present in the experimental observation while NS fails to do the same.}

Finally, the result provide a better understanding and characterization of the relationship between the stress tensor and the nonequilibrium processes. More accurate prediction of the material parameters obtained from the first order approximation to Boltzmann-Curtiss equation could lead to further significant improvements in the capability of continuum solvers to model nonequilibrium hypersonic flows.

\begin{acknowledgment}
 Authors would like especially express gratitude to Dr. Eswar Josyula for insightful discussions and DSMC data for validations.
\end{acknowledgment}

\section*{Funding}
This material is based upon work supported by the Air Force Office of Scientific Research under award number FA9550-17-1-0154.



%
% ...

\newpage
\bibliographystyle{asmems4}

\bibliography{asme2e}







\newpage
\section*{Figures}

\begin{figure}[H]
  \caption{Boltzmann-Curtiss distribution function of the velocity and gyration of the particles} \label{fig1}
\end{figure}


\begin{figure}[H]
  \caption{Comparison of argon's reciprocal shock thickness} \label{fig2}
\end{figure}

\begin{figure}[H]
\caption{Shock structure of argon at Mach 3.38. (a) Density profile; (b) Reciprocal shock thickness} \label{fig3}
\end{figure}


 \begin{figure}[H]
\caption{Argon normal stresses and heat flux, at Mach 1.2 (a $\&$ c) and Mach 8 (b $\&$ d). Figures (a) and (b) show the normal stresses with color black for $t_{xx}$, \textcolor{black}{blue: $t_{yy}$, and figures (c) and (d) show the heat flux}
Normal stresses of argon,; }  \label{Argon_stress_q}
\end{figure}

\begin{figure}[H]
  \caption{Comparison of nitrogen's reciprocal shock thickness} \label{nitrogen_thickness}
\end{figure}

 \begin{figure}[H]
\caption{Nitrogen density and temperature profile (black: translation, \textcolor{black} {blue: rotation}) at Mach 1.2 (a $\&$ b), Mach 5 (c $\&$ d), and Mach 10 (e $\&$ f). }  \label{N2_All}
\end{figure}


\end{document}